\documentclass[aps,prl,twocolumn,superscriptaddress,showpacs]{revtex4}
\usepackage{graphicx}
\begin{document}

\title{High-Mobility Few-Layer Graphene Field Effect Transistors Fabricated on Epitaxial Ferroelectric Gate Oxides}

\author{X. Hong}
\affiliation{Department of Physics, The Pennsylvania State
University, University Park, PA 16802}

\author{A. Posadas}
\affiliation{Department of Applied Physics, Yale University, New
Haven, CT 06520}
\author{K. Zou}
\affiliation{Department of Physics, The Pennsylvania State
University, University Park, PA 16802}
\author{C. H. Ahn}
\affiliation{Department of Applied Physics, Yale University, New
Haven, CT 06520}
\author{J. Zhu}
\affiliation{Department of Physics, The Pennsylvania State
University, University Park, PA 16802}

\date{\today}

\begin{abstract}
The carrier mobility $\mu$ of few-layer graphene (FLG) field-effect
transistors increases ten-fold when the SiO$_2$ substrate is
replaced by single-crystal epitaxial Pb(Zr$_{0.2}$Ti$_{0.8}$)O$_3$
(PZT). In the electron-only regime of the FLG, $\mu$ reaches
7$\times10^4$ cm$^2$/Vs at 300K for $n$ = $2.4\times10^{12}$/cm$^2$, 70\%
of the intrinsic limit set by longitudinal acoustic (LA) phonons; it increases to
1.4$\times10^5$cm$^2$/Vs at low temperature. The
temperature-dependent resistivity $\rho(T)$ reveals a clear
signature of LA phonon scattering, yielding a deformation potential
$D$ = 7.8$\pm$0.5 eV.

\end{abstract}
 \maketitle

 Recent calculations show that the intrinsic mobility of graphene, set by
 longitudinal acoustic (LA) phonon scattering, can reach $\sim$10$^5$ cm$^2$/Vs at
 room temperature \cite{hwang2008}. However, extrinsic scattering sources, many of which
 arise from the surface morphology, chemistry, structural and electronic
 properties of the widely used SiO$_2$ substrate, limit the mobility to the
 current range of $2\times 10^3 - 2\times 10^4$ cm$^2$/Vs \cite{ando2006,geim2007,nomura2007,hwang2007,tan2007,morozov2008,katsnelson2008,fratini2008,hwang2008,chen2008}
. Increasing the mobility beyond these extrinsic limits is one of
the central challenges of the graphene community. Recently, two
groups have reported a significant improvement in the mobility of
suspended graphene after current-heating annealing
\cite{bolotin2008a,Du2008}. A more device-friendly solution involves
placing graphene on a different substrate. Studies on several
alternatives have been reported recently although these substrates
result in mobilities comparable to that on SiO$_2$ \cite{Geim2008}.

In this letter we report significant carrier mobility improvement in
few-layer graphene (FLG) field effect transistors (FETs) fabricated
with single-crystal epitaxial Pb(Zr$_{0.2}$Ti$_{0.8}$)O$_3$ (PZT)
films as the gate oxide. At 300 K, PZT-gated FLG exhibits a mobility
$\mu \sim 7\times10^4$ cm$^2$/Vs at a density of $n$ =
$2.4\times10^{12}$/cm$^2$, reaching 70\% of the intrinsic limit set
by LA phonons. We observe a clear signature of LA phonon scattering
in the temperature dependence of resistivity $\rho(T)$. The
PZT-gated FLG shows a residual resistivity $\rho_0$ at low
temperature approximately an order of magnitude lower than that of
SiO$_2$-gated single and few-layer graphene. This low $\rho_0$
corresponds to a mobility of $1.4\times10^5$ cm$^2$/Vs and a long
carrier mean free path of 2 $\mu$m at $n=2.4\times10^{12}$/cm$^2$.
Our results open up a promising route into realizing graphene's full
scientific and technological potential
\cite{geim2007,dassarma2007,avouris2007}.

\begin{figure}[htbp]

\includegraphics[angle=0,width=3.3in]{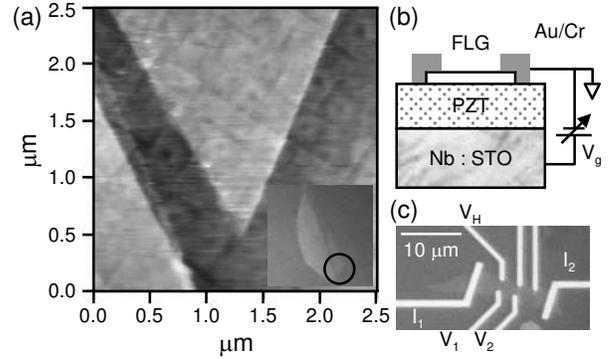}
\vspace{-0.1in} \caption[]{(a) AFM contact mode image of a 2.4 nm
FLG flake (center) on a 400 nm PZT film. Inset: optical image of the
whole flake with the area in (a) circled. The PZT surface shows
smooth terraces separated by $a$-axis lines, with a root-mean-square
(RMS) surface roughness of 3-4 {\AA} over a 1 $\mu$m$^2$ area. FLG
has a roughness of 2-3 \AA. (b) Device schematics. (c) Hall bar
configuration of a FLG-FET with current (I$_1$, I$_2$) and voltage
electrodes for resistance (V$_1$ and V$_2$) and Hall (V$_1$ and
V$_H$) measurements. We determine the thickness of this FLG to be
(2.4$\pm$0.3) nm based on its optical transparency. \label{AFM}}
\end{figure}

For our FETs we work with 400 nm PZT films that are epitaxially grown on Nb-doped
single-crystal SrTiO$_3$ (STO) substrates via radio-frequency
magnetron sputtering \cite{hong2003}. X-ray diffraction and atomic
force microscopy (AFM) measurements show that these films have high
crystalline and surface quality (Fig.~\ref{AFM}(a))
\cite{supporting}.

FLG flakes are mechanically exfoliated on PZT from Kish graphite
(Toshiba Ceramics Co.) followed by optical identification and AFM
characterization (Fig.~\ref{AFM}(a)) \cite{supporting}. Multiple
electrodes (30 nm Au/3 nm Cr) in the Hall-bar configuration are
patterned onto regularly shaped flakes using e-beam lithography and
metal evaporation. The Nb-doped STO substrate serves as the backgate
to which a bias voltage $V_g$ is applied to tune the carrier density
of the FLG (Figs.~\ref{AFM}(b) and (c)). Results reported here are
collected from 3 FETs fabricated on the same PZT substrate and one
FET on a SiO$_2$ substrate.

\vspace{-0.1in}

\begin{figure}[htbp]
\includegraphics[angle=0,width=3.1in]{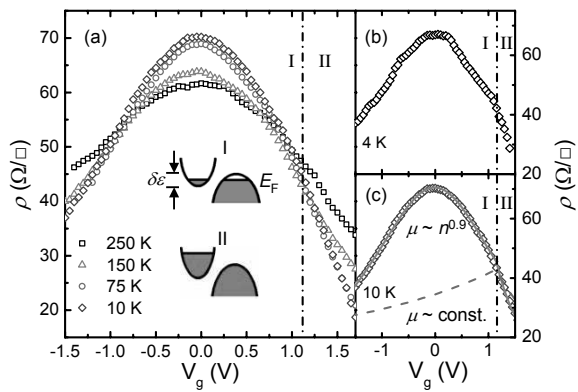}
\vspace{-0.1in}
 \caption[]{(a) $\rho$($V_g$) at selected
temperatures taken on the device shown in Fig.~\ref{AFM}(c). Inset:
schematics of the band structure of FLG of this thickness. (b)
$\rho$($V_g$) at 4 K. The kink at $V^{T}_{g}$ = 1.1 V (dash-dotted
line) marks the boundary between regimes I and II. (c) $\rho$($V_g$)
at 10 K (open symbols) with a fitting curve (solid line) combining
Eqs.~\ref{eq1} and \ref{eq3} with $\beta$ = 0.9 and $r$ = 0.6. The
dashed line is calculated from Eq.~\ref{eq1} assuming a
density-independent mobility $\mu_e=\mu_h=1\times 10^{5}$cm$^2$/Vs.
\label{resistivity}}
\end{figure}

Resistivity and Hall measurements were performed in a $^4$He cryostat
with a base temperature of 1.4 K, equipped with a superconducting
magnet. Standard low frequency (47Hz) lock-in techniques are used
with excitation currents ranging from 50 to 200 nA. In
Fig.~\ref{resistivity}, we show the sheet resistivity $\rho$ of a
FLG device ($\sim$2.4 nm or 7 layers, Fig.~\ref{AFM}(c)) as a
function of $V_g$ at temperatures 4 K $< T <$ 300 K. $\rho$($V_g$)
displays a broad maximum at the charge neutrality point. Curves
below 300 K are shifted to align the $\rho$($V_g$) maximum at $V_g$
= 0 V~\cite{shift}. FLG of this thickness behaves as a
two-dimensional (2D) semimetal, where the low-energy bands for
electrons and holes are parabolic and overlap slightly
\cite{novoselov2004} (inset of Fig.~\ref{resistivity}(a)).
 The carrier density in the FLG is
controlled by $V_g$ through $n_e$-$n_h$ = $\alpha$$V_g$, where
$\alpha$ is the charge injection rate of the backgate. In the
band-overlap regime (regime I in Fig.~\ref{resistivity}(a) inset),
both electrons and holes contribute to conduction:
\begin{equation}
\label{eq1} \frac{1}{\rho}=e(n_e\mu_e+n_h\mu_h)
\end{equation}
At sufficiently large $\mid$V$_g\mid$, the system becomes a pure 2D
electron (regime II in Fig.~\ref{resistivity}(a) inset) or hole (not
shown) gas \cite{novoselov2004}. There, the resistivity and the Hall
coefficient $R_H$ are given by:
\begin{equation}
\label{eq2} \frac{1}{\rho}=en_{e,h}\mu_{e,h};\
R_H=\frac{1}{en_{e,h}};\ n_{e,h}=\alpha V_g
\end{equation}
We measure $R_H$ in the hole-only regime of two devices and
determine $\alpha=1.35\times 10^{12}$ cm$^{-2}$/V. Using a parallel-plate capacitor model, we
extract a dielectric constant $\kappa\sim 100$ for our PZT films.
This value is confirmed by independent low-frequency capacitance
measurements (see Ref.~\cite{supporting} for details). Compared with
$\kappa = 3.9$ in SiO$_2$, PZT gate oxides are extremely efficient
in injecting carriers into graphene, as well as screening charged
impurities in its vicinity.

It is clear from Eqs.~\ref{eq1} and \ref{eq2} that the slope of
$\rho$($V_g$) changes at a threshold $V^{T}_{g}$, where the system
transitions from a two-carrier to a single-carrier regime.
Such a slope change has been reported in Bismuth nanowires, which is
another two-carrier system \cite{boukai2006}. Indeed, a kink at $V^{T}_{g}$ = 1.1 V is clearly visible in
$\rho$($V_g$) at low temperature (Fig.~\ref{resistivity}(b)), where
$n_e = 1.5\times 10^{12}$/cm$^{2}$ and $n_h$ = 0. Modeling the FLG
in regime I with one electron and one hole band and using the
effective mass values determined in Ref.~\cite{novoselov2004} for
this thickness ($m^{*}_e$ = 0.06 $m_0$ and $m^{*}_h$ = 0.10 $m_0$),
we estimate the electron and hole densities at the charge neutrality
point to be $n^{0}_{e}$ = $n^{0}_{h}\sim 9\times 10^{11}$/cm$^{2}$.
This corresponds to an overlap between the electron and hole bands
of $\sim$30 meV (see Ref.~\cite{supporting} for more discussions).
These estimates are in good agreement with results obtained using
methods described previously \cite{novoselov2004} and band structure
calculations of FLG of this thickness \cite{partoens2006}. These
studies also suggest that FLG in this thickness range may have more
than one hole band \cite{novoselov2004,partoens2006}. We emphasize
that the central results of the present study are taken in the
electron-only regime described by Eq.~\ref{eq2}, and do not rely on
the accurate knowledge of the band structure in the two-carrier or
hole-only regimes.

In single and few-layer graphene fabricated on SiO$_2$ substrates,
the mobility has been found to be roughly density-independent and
explained in terms of long-range charged impurity scattering
\cite{ando2006,geim2007,nomura2007,hwang2007,tan2007,morozov2008,novoselov2004}.
$\rho$($V_g$) calculated using this assumption and also using Eq.~\ref{eq1} is
plotted in Fig.~\ref{resistivity}(c) (dashed curve). Clearly the $n$-independent assumption does not describe our data (open symbols) in the band-overlap regime (I).
Instead, a density-dependent mobility  $\mu_{e,h}\sim
n^{\beta}_{e,h}$ produces an excellent fit to the data within the
entire regime. The power-law functional form is motivated by
measurements in regime II, shown later. The solid line in
Fig.~\ref{resistivity}(c) shows such fitting with mobilities
determined by:
\begin{equation}
 \label{eq3}
 \mu_{e}(n_e)=cn^{\beta}_{e};\ \mu_{h}(n_h)=crn^{\beta}_{h}
\end{equation}
\begin{figure}[htbp]
\includegraphics[angle=0,width=2.8in]{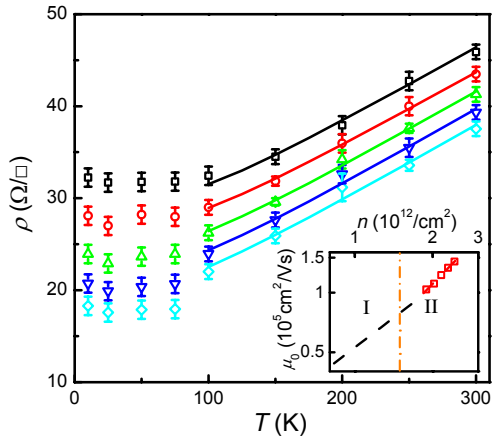}
\vspace{-0.1in} \caption[]{(Color online) $\rho$($T$) at electron
densities of (from top to bottom) $n$ = 1.89, 2.02, 2.16, 2.30 and
2.43$\times 10^{12}$/cm$^{2}$. The solid lines are fittings to
Eq.~\ref{eq4} for $T >$ 100 K, with the corrections due to a
non-degenerate Fermi gas included. Inset: Low-$T$ residual mobility
$\mu_{0}(n)$ in a double-log plot. Open squares are data taken in
regime II. The dashed line plots the fitting (Eq.~\ref{eq3},
electrons) obtained in regime I. \label{phonon}}
\end{figure}
where we require $\mu_e$ and $\mu_h$ to have a power-law dependence
on $n_e$ and $n_h$ respectively with the same exponent $\beta$ but
scales by a factor $r$. We obtain  $\beta$ = 0.9 from the fit in
Fig.~\ref{resistivity}(c). The constant $c$ is determined by
matching a measured data point $\mu_e = 1.0\times10^{5}$cm$^{2}$/Vs
at the electron density of $n$ = 1.75$\times10^{12}$/cm$^2$ in
regime II. The approximate symmetric $V_g$-dependence $\rho$($V_g$)
displayed for both carriers in regime I, together with $\beta \sim
1$, implies that $r=\frac{\mu_{h}(n_h)}{\mu_e(n_e)}\sim m^{*}_e/
m^{*}_h \sim 0.6$. In addition to $\rho(V_g)$, the above fitting
parameters also produce excellent agreement with $R_H$ data in
regime I (see Ref.~\cite{supporting} for details). The origin of the
mobility asymmetry is unclear to us at the moment although we note
that such phenomenon also occurs in some SiO$_2$-gated graphene
devices~\cite{bolotin2008a,Chen2008a}. The $n$-dependence of $\mu$,
on the other hand, distinguishes our device from SiO$_2$-gated
graphene and FLG \cite{geim2007,tan2007,morozov2008,novoselov2004}.
This $n$-dependence may originate from the overlapped bands, or
suggest different scattering mechanisms. Results from the
electron-only regime shown later seem to support the second
explanation.

The above analysis provides an approximate scenario of transport in
the band-overlap regime of the FLG. Below we present and analyze the
central results of our work, derived from data taken in regime (II)
of the FLG ($V^{T}_{g}>$ 1.1 V), where the FLG behaves as a
single-carrier, one band, two-dimensional electron gas described by
Eq.~\ref{eq2}. Figure~\ref{phonon} plots $\rho(T)$ extracted from data
shown in Fig.~\ref{resistivity}(a) at five electron densities ranging from
1.9$\times 10^{12}$/cm$^2$ (at $V_g$ = 1.4 V) to 2.4$\times
10^{12}$/cm$^2$ (at $V_g$ = 1.8 V), well into regime II. At a fixed $n$,
$\rho(T)$ follows a linear $T$-dependence between 100 K and 300 K
and quickly saturates to a non-zero residual value $\rho_{0}(n)$ at
lower $T$. This linear $T$-dependence, its temperature range, and
the magnitude of the resistivity change strongly point to scattering
between electrons and LA phonons in graphene. This phonon mode,
nearly identical in graphene and graphite, has been calculated
\cite{hwang2008} and experimentally studied \cite{chen2008} recently
in graphene on SiO$_2$. However, the combination of a large
$\rho_{0}$ and the onset of another scattering mechanism at 150 K in
SiO$_2$-gated graphene makes it difficult to extract the LA phonon
contribution unambiguously in those systems \cite{chen2008}.
\begin{figure}[htbp]
\includegraphics[angle=0,width=2.6in]{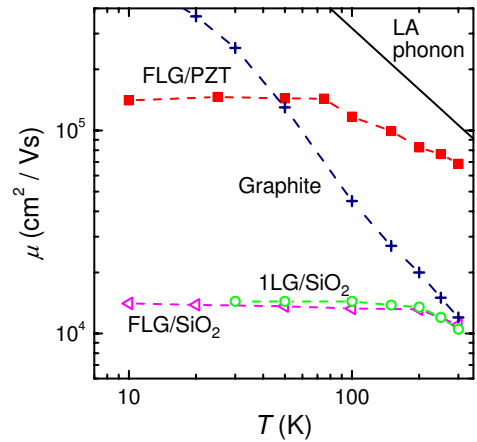}
\vspace{-0.1in}\caption[]{(Color online) Comparison of $\mu$($T$) in
various graphitic materials. Solid squares: PZT-gated FLG shown in
Fig.~\ref{phonon} at $n$ = 2.4$\times 10^{12}/cm^{2}$. Open
triangles: a SiO$_2$-gated FLG of the same thickness and
density\cite{supporting}. Open circles: single-layer graphene on
SiO$_2$ reported in Ref.~\cite{chen2008}. Crosses: mobility of bulk
graphite from Ref.~\cite{ono1966}. Solid line: LA phonon-limited
mobility calculated from Eq.~\ref{eq4}. \label{comparison}}
\end{figure}
In our devices, a small $\rho_0$ enables us to clearly observe the
predicted linear $T$-dependence at $T
> T_{BG}$, where $T_{BG}=\frac{2\hbar k_{F}v_{ph}}{k_B}\approx 80 K$
is the Bloch-Gruneisen temperature at $n$ = 2$\times
10^{12}$/cm$^2$, using a sound velocity $v_{ph}$ = 2.1$\times 10^6$
cm/s for LA phonons in graphene and $k_{F}=\sqrt{\pi n}$ for the
Fermi wave vector of the 2D electron gas. At $T > T_{BG}$, the
contribution to the resistivity from LA phonons is given by:
\begin{equation}
 \label{eq4}
\rho_{ph}(T,
n)=\frac{m^{*}_{e}}{ne^2}\langle\frac{1}{\tau}\rangle=\frac{1}{n}\frac{(m^{*}_{e})^{2}D^{2}k_{B}T}{4\hbar^{3}e^{2}\rho_{m}v^{2}_{ph}}
 \end{equation}
where we have modified the derivation in Ref.~\cite{hwang2008} to
account for massive electrons in FLG. $D$ is the unscreened acoustic
deformation potential \cite{supporting} and
$\rho_m$ = 6.5$\times 10^{-7}$ kg/m$^2$ is the areal mass density of
graphene. The correction due to a non-degenerate Fermi gas is less
than a few percent in our density and temperature range and is
neglected in Eq.~\ref{eq4}.

Solid lines in Fig.~\ref{phonon} show fittings at different
densities for $T >$ 100 K, where the slopes range from 83 to 87
m$\Omega$/K and lead to $D$ = 7.8$\pm$0.5 eV in graphene. This
result falls within the range of reported values in the literature
of 1 - 30 eV
\cite{chen2008,ono1966,pietronero1980,yang1999,woods2000,du2005} and
agrees very well with tight-binding calculations producing $D$ $\sim
3\gamma$ , where $\gamma \sim$ 3 eV is the nearest-neighbor hopping
matrix \cite{yang1999}. We do not observe evidence of super-linear
$T$-dependences reported in graphene on SiO$_2$
\cite{morozov2008,chen2008} that are attributed to remote substrate
\cite{fratini2008,chen2008} or inter-ripple flexural phonons
\cite{morozov2008}. We speculate that a higher
stiffness and a larger average carrier-substrate separation in FLG
may be responsible for suppressing scatterings from these two types
of phonons.

The small residual resistivity $\rho_{0}$ in PZT-gated FLG leads to
mobility $\mu_{0}$ in excess of $1\times 10^{5}$ cm$^2$/Vs at low
$T$. Since both FLG and single layer graphene are subject to similar
scattering mechanisms, a comparison between $\mu$ of PZT-gated FLG,
SiO$_2$-gated FLG and SiO$_2$-gated graphene highlights the
important role played by the substrate. Such comparison is shown in
Fig.~\ref{comparison}, where we compare $\mu$($T$) obtained from two
2.4 nm thick FLG (one on PZT, one on SiO$_2$\cite{supporting}),
graphene on SiO$_2$ from Ref.~\cite{chen2008}, bulk graphite from
Ref.~\cite{ono1966} and the intrinsic LA phonon-limited mobility
calculated from Eq.~\ref{eq4}, using $D$ = 8 eV. At a density of $n$
= 2.4$\times 10^{12}$/cm$^2$, the PZT-gated device shows $\mu\sim
7\times 10^{4}$ cm$^2$/Vs at room temperature,
 70$\%$ of the intrinsic phonon mobility of $\sim 1\times 10^{5}$ cm$^2$/Vs.
At low $T$, $\mu$ increases to $1.4\times 10^{5}$ cm$^2$/Vs,
corresponding to a long mean free path of 2 $\mu$m. A second device
($\sim$5nm thick, not shown) on the same PZT substrate exhibits
mobilities of $7.5\times 10^{4}$ cm$^2$/Vs at room temperature and
$1.5\times 10^{5}$ cm$^2$/Vs at low temperature. These values
represent an approximately ten-fold increase over those of our
SiO$_2$-gated FLG as well as single and few-layer graphene reported
in the literature, where $\mu$ ranges $2\times 10^3 - 2\times 10^4$
cm$^2$/Vs with weak or no temperature dependence
\cite{tan2007,morozov2008,chen2008,novoselov2004}. This remarkable
mobility improvement clearly demonstrates the advantage of the PZT
substrate over SiO$_2$ towards fabricating graphene-based
high-quality 2D systems.

The low-temperature residual mobility $\mu_{0}(n)$ in PZT-gated FLG
exhibits a density dependence best described by $\mu_{0}(n)\sim
n^{1.3}$ for 1.9$\times 10^{12}/$cm$^2$ $< n <$ 2.4 $\times
10^{12}/$cm$^2$. In the inset of Fig.~\ref{phonon}, we show
$\mu_{0}(n)$ data in this range together with the fitting obtained
in regime I: $\mu_{0}(n)\sim n^{0.9}$. This $n$-dependence of $\mu$
is in contrast to the SiO$_2$-gated graphene, where the scattering
due to Coulomb impurities leads to a very weak $n$-dependence in a
comparable density range, suggesting that different scattering
mechanisms are at work
\cite{ando2006,geim2007,nomura2007,hwang2007,tan2007,morozov2008,novoselov2004,price1984,pfeiffer1989}.

It has been shown in suspended graphene that a significant
improvement in $\mu$ is only achieved after current annealing, which
highlights the important role played by interfacial adsorbates
\cite{bolotin2008a}, among other proposed scattering mechanisms at
low temperature
\cite{morozov2008,katsnelson2008,fratini2008,chen2008,martin2008}.
Our PZT substrates possess a large spontaneous polarization $P$
pointing into the surface \cite{supporting}. The absence of free
carriers in ungated FLG devices indicates that this polarization is
almost completely screened by a high-density layer of surface
adsorbates prior to exfoliation. Screening adsorbates may come from
free ions, atoms and molecules in the ambient and OH$^-$ and H$^+$
produced by the dissociation of H$_{2}$O on PZT surface
\cite{peter2004,fong2006,henderson2002}. Despite their high density,
our data suggest that the scattering from interfacial adsorbates is
much weaker than in SiO$_2$-gated devices. We attribute this
remarkable phenomenon to the strong screening of PZT and speculate
that some degree of ordering in the adsorbate layer may also play a
role in reducing the scattering.

In conclusion, we have demonstrated a significant performance
improvement in few-layer graphene FETs by using the crystalline
ferroelectric gate oxide PZT. This approach has led us to the
observation of the highest reported mobility to date in unsuspended
single and few-layer graphene devices. This result opens up a new
route for realizing high-speed electronic devices and exploring
novel 2D physics in graphene.

\begin{acknowledgments}
We are grateful for helpful discussions with V. Crespi, P. Eklund,
V. Henrich, J. Hoffman, J. Jain, P. Lammert, G. Mahan, J. Reiner, H.
Stormer and F. Walker. Work at Penn State is supported by NSF NIRT
ECS-0609243 and NSF CAREER DMR-0748604. Fabrication of samples at
Yale is supported by NSF MRSEC DMR-0520495, NSF DMR-0705799, ONR,
and SRC. The authors also acknowledge use of facilities at the PSU
site of NSF NNIN.
\end{acknowledgments}




\end{document}